\documentclass[12pt,preprint]{aastex}

\slugcomment{Submitted to ApJ}

\shorttitle{Mass loss history of IRC+10420}
\shortauthors{Dinh-V-Trung et al.}

\newcommand{\kms}{\mbox{km\,s$^{-1}$}}

\begin{document}

\title{Probing the mass loss history of the yellow hypergiant IRC+10420}

\author{Dinh-V-Trung\altaffilmark{1}, S\'ebastien Muller, Jeremy Lim}
\altaffiltext{1}{on leave from Center for Quantum Electronics, Institute of Physics,
Vietnamese Academy of Science and Technology,
10 DaoTan, ThuLe, BaDinh, Hanoi, Vietnam}
\email{trung@asiaa.sinica.edu.tw, muller@asiaa.sinica.edu.tw, jlim@asiaa.sinica.edu.tw}
\affil{Institute of Astronomy and Astrophysics, Academia Sinica\\ 
P.O Box 23-141, Taipei 106, Taiwan.}
\and
\author{Sun Kwok}
\affil{Department of Physics, Faculty of Science, University of Hong Kong}
\email{sunkwok@hku.hk}
\and 
\author{C. Muthu}
\affil{Aryabhatta Research Institute of Observational Sciences (ARIES), \\
Manora Peak, Nainital, India}
\email{muthu@aries.ernet.in}
\begin{abstract}
We have used the sub-millimeter array to image the molecular envelope
around IRC+10420. Our observations reveal a large and clumpy expanding envelope around the star. 
The molecular envelope shows a clear asymmetry in $^{12}$CO J=2--1 emission in the South-West direction. 
The elongation of the envelope is found even more pronounced in
the emission of $^{13}$CO J=2--1 and SO J$_{\rm K}$=6$_5$--5$_4$. A small positional velocity gradient across
velocity channels is
seen in these lines, suggesting the presence of a weak bipolar outflow in the envelope of
IRC+10420. 
In the higher resolution $^{12}$CO J=2--1 map, we find that 
the envelope has two components: (1) an inner shell (shell I) located between radius
of about 1"-2"; (2) an outer shell (shell II) located between 3" to 6" in radius. These shells represent two previous 
mass-loss episodes from IRC+10420. We attempt to 
derive in self-consistent manner the physical conditions inside the envelope
by modelling the dust properties, and the heating and cooling of molecular gas. We estimate
a mass loss rate of $\sim$9 10$^{-4}$ M$_\odot$ yr$^{-1}$ for shell I and 7 10$^{-4}$ M$_\odot$ yr$^{-1}$ for
shell II. The gas temperature is found to be  unusually high in IRC+10420 in comparison with other oxygen-rich envelopes.
The elevated gas temperature is mainly due to higher heating rate, which results from the large luminosity of the central star.
We also derive an isotopic ratio $^{12}$C/$^{13}$C = 6.
\end{abstract} 

\section{Introduction}
Yellow hypergiants are rare objects which are post-red supergiants rapidly evolving in
blue-ward loops in the Hertzsprung-Russell diagram (de Jager 1998).  
A few of such objects are found to have prodigous mass loss, leading to the formation of thick
circumstellar envelopes. The study of the circumstellar envelope around them is very important
to provide insight into the mass loss process from these massive stars and the evolution of
stars themselves.

The star IRC+10420 is classified as yellow hypergiant (de Jager 1998). Based on
spectroscopic monitoring, the star initially classified as spectral type
F8 I$^{+}_{\rm a}$ in 1973 (Humphreys et al. 1973), has evolved into spectral type A in 1990 (Oudmaijer
et al. 1996, Klochkova et al. 1997). Strong emission lines such as the Balmer series and 
Ca II triplets are seen in its optical spectrum (Oudmaijer 1998). Broad wings seen in H$\alpha$
and Ca II triplets suggest a large outflow velocity in a massive wind close to the stellar 
surface. High resolution optical images of IRC+10420 (Humphreys et al. 1997) revealed a complex circumstellar environment
with many features such as knots, arcs or loops. Lacking kinematic information, however, the real three
dimensional structure of the envelope can not be inferred. Recently, through integral-field spectroscopy
of H$\alpha$ emission line, Davies et al. (2007) show strong evidence for axi-symmetry in the
circumstellar envelope.

The molecular envelope around IRC+10420 is very massive and is a rich source of molecular lines (Quintana-Lacaci et al. 2007).
The mass loss rate estimated from CO observations is at the upper end seen toward evolved stars, 
up to a few times 10$^{-4}$ M$_\odot$ yr$^{-1}$ (Knapp \& Morris 1985, Oudmaijer et al. 1996).
High denisty tracers such as ammonia (NH$_3$) and HCN lines have been detected \citep{menten95}.
Molecules typical of oxygen-rich envelope such as SO and SO$_2$ (Sahai \& Wannier 1992,
Omont et al. 1993) are known to be present. 
IRC+10420 is also a strong OH maser source, exhibiting both
the main lines (1665 MHZ and 1667 MHz) maser and the satelite line (1612 MHz) maser. By comparison
with typical OH/IR stars, IRC+10420 has the warmest envelope harboring OH masers (Humphreys et al. 1997).
High angular resolution imaging of the OH masers by \cite{nedoluha92} shows that the maser emissions are distributed
in an oblate expanding shell measuring 1\arcsec.5 in radius. Recently, Castro-Carrizo et al. (2007) presented
high angular resolution ($\sim$3\arcsec\, and 1\arcsec.4) mapping of the envelope around IRC+10420 in $^{12}$CO J=1--0 and J=2--1 lines.
Their observations reveal an expanding but complex molecular envelope with an inner
cavity around the central star, implying a recent dramatic decrease in the mass loss from IRC+10420. In addition,
the shell shows some elongation in the North-East to South-West direction, suggesting a non-spherical symmetry in the
mass loss process. 
From excitation analysis of $^{12}$CO lines, Castro-Carrizo et al. (2007) show the 
presence of two separate shells with mass loss in the range 1.2 10$^{-4}$ to
3 10$^{-4}$M$_\odot$ yr$^{-1}$. Interestingly, the gas temperature in shells derived from their model fitting
is unusually high, 100 K even at the large radial distance of $\sim$2 10$^{17}$ cm.

The distance to IRC+10420 is still uncertain. From analysis of photometric and polarimetric data,
Jones et al. (1993) conclude that IRC+10420 is located at a large distance, between 4 to 6 kpc.
Here we adopt a distance to IRC+10420 of 5 kpc, similar to that adopted in Castro-Carrizo et al. (2007).

In this paper we present high angular resolution observations of the molecular envelope around IRC+10420.
We also perform detailed modelling to derive in a self-consistent manner the dust properties and 
molecular gas temperature profile in the envelope. 

\begin{table}[h] \begin{center} \begin{tabular}{lll}
\multicolumn{3}{l}{Table 1: Basic data of IRC+10420 and its molecular envelope} \\
\hline
\hline
R.A. (J2000)  & 19$^{\rm h}$26$^{\rm m}$48$\fs$03 & (1) \\  
Dec. (J2000)   & 11$\degr$21$\arcmin$16$\farcs$7   & (1)\\ 
Distance      &  5 kpc                & (2) \\
Angular scale & 1$\arcsec$ $\sim$  5000 AU $\sim$  7.5 10$^{16}$ cm  & \\
Luminosity L$_*$ &  $\sim$ 6 10$^5$ L$_\odot$ & (2)  \\
Spectral Type & mid-A type  & (3,4) \\
Effective Temperature T$_{\rm eff}$ & 7000 K & (3) \\
Stellar radius & 1.66 Au \\
Systemic velocity & V$_{\rm LSR}$ = 74 kms$^{-1}$  & (3) \\
Expansion velocity & V$_{\rm exp}$ = 40 kms$^{-1}$ & (3)\\
Mass loss rate  $\dot{M}$ & $\sim$ 6.5 10$^{-4}$ M$_\odot$ yr$^{-1}$ & (5) \\
\hline
\end{tabular}
\tablerefs{ (1) SIMBAD database;
(2) Jones et al. (1993);
(3) Oudmaijer et al. (1996);
(4) Klochkova et al. 1997;
(5) Knapp \& Morris (1985)}
\end{center}
\end{table}

\section{Observations}

We have observed IRC+10420 in the 1.3mm band with the SubMillimeter Array (SMA)
at two different epochs and in two different configurations as summarized in Table 2. The first observation
was carried out on 2004 June 21$^{st}$, with 8 antennas in an extended configuration, giving projected
baselines from 27 to 226 m. The weather was good during the observation with zenith atmospheric opacity of
 $\tau_{\rm 225GHz} \sim 0.2$ at 225 GHz and the system temperatures ranged between 200 and 600 K. 
The second observation was carried on 2005 July 2$^{nd}$, with 7 antennas in a compact configuration, giving projected baselines
in the range from 11 to 69 m. The weather was very good with $\tau_{\rm 225GHz} \sim 0.1$ and the system temperatures ranged
between 100 and 200 K.

In each observation, the correlator of the SMA was configured in the normal mode with two 2-GHz bandwidth windows
separated by 10 GHz. This allowed us to cover simultaneously the $^{12}$CO J=2-1 line in the upper sideband (USB)
the $^{13}$CO J=2-1 and SO 6$_{\rm 5}$-5$_{\rm 4}$ lines in the lower sideband (LSB). 
The bandpass of individual antennas was calibrated using the quasar 0423-013 for the observation taken
in 2004 and Uranus for observation taken in 2005.
To improve the S/N of the data,
we smoothed data from the instrumental spectral
resolution of 0.8125 MHz to 3.25 MHz ({\em i.e.} $\sim$ 4 km~s$^{-1}$). For comparison, the 
observed lines have velocity widths of ($\sim 80$ km~s$^{-1}$).
The nearby quasars 1751+096 and
1925+211 were observed at regular intervals to correct for gain variations of the antennas due to atmospheric fluctuations. 
The data reduction and calibration were done
under MIR/IDL\footnote{http://cfa-www.harvard.edu/$\sim$cqi/mircook.html}. After the gain and bandpass calibrations, the
visibilities data from both tracks were combined together.
Imaging and deconvolution were performed under GILDAS\footnote{http://www.iram.fr/IRAMFR/GILDAS/} package.

The 1.3 mm (225 GHz) continuum emission map was derived by averaging the line-free channels from
both LSB and USB, resulting in a total bandwidth of 3.4 GHz. For the line channel maps we subtract
the continuum visibilities of the relevant
sideband from the line emission visibilites. In Table~\ref{tabobs} we summarize the observations of IRC+10420 carried out by the SMA.

For the sake of presentation clearity, the center of all the maps were shifted in the visibility domain so that
the phase reference matches the position of the continuum peak emission,
{\em i.e.} R.A. = 19$^{\rm h}$26$^{\rm m}$48$\fs$09 and Dec. = 11$\degr$21$\arcmin$16$\farcs$75,
which also corresponds to the center of the SiO shell detected by \citet{cas01}.
We note that the lack of interferometric baselines shorter than 11 m makes our observations blind to
extended structures of size $\ge 25\arcsec$. In order to estimate the amount of flux recovered by the SMA, we convolved
the channel maps of $^{12}$CO J=2--1 and $^{13}$CO J=2--1 lines to the spatial resolution of 19\arcsec.7 and 12\arcsec,
respectively. We show in Figure 2 the spectra taken from the these maps and the single dish observations taken by
Kemper et al. (2003) and Bujarrabal et al. (2001) at the same angular resolutions. We estimate that the SMA 
recovered more than 60\% of the flux in $^{12}$CO J=2--1 line and almost all the flux in $^{13}$CO J=2--1 line.
\begin{table}[h] \label{tabobs} \begin{center} \begin{tabular}{lllcccc}
\multicolumn{7}{l}{Table 2: SMA line observations towards IRC+10420.} \\
\hline
\hline
Line & Rest Freq & Config. & Clean beam & P.A.       & $\Delta$V & 1$\sigma$ rms \\
                         & (GHz) &   & (FWHM, arcsec)         & (deg)      & (km~s$^{-1}$)  & (mJy/beam)    \\
\hline

$^{12}$CO(2-1)           & 230.538 & Comp  & 3.44 x 3.12      & 46$^\circ$ & 4.2                 & 60 \\
                         &         & Comp+Ext & 1.44 x 1.12      & 46$^\circ$ & 4.2                 & 80 \\

$^{13}$CO(2-1)           & 220.399 & Comp & 3.52 x 3.21      & 63$^\circ$ & 4.4                 & 50 \\
SO(6$_5$-5$_4$)          & 219.949 & Comp & 3.51 x 3.19      & 63$^\circ$ & 4.4                 & 50 \\
\hline
\end{tabular} \end{center} \end{table}

\section{Results}
\subsection{1.3mm continuum emission}

As shown in Figure 1, the 1.3mm continuum emission from IRC+10420 was well detected in our SMA data. 
The emission is not resolved and has a flux density $\sim$ 45 mJy. A previous single-dish observation by
\citet{walmsley91} also at 1.3mm
gives a flux of 101 mJy. That suggests that the dust emission is extended, with 
about 55\% of the continuum flux resolved out by the interferometer.
Assuming the central star to have a black body spectrum with a temperature of T${\rm eff}$ = 7000 K, it
contributes only a negligible amount of the flux at 1.3 mm of S$_*$ $\sim$ 0.1 mJy.
Thus most of the 1.3mm continuum emission
is produced by the warm dust in the envelope around IRC+10420.

\subsection{$^{12}$CO J=$2-$1 emission}
In Figure. 3 we show the channel maps of the $^{12}$CO J=2$-$1 emission
obtained with the compact configuration of the SMA. At our angular resolution of $\sim$3\arcsec, the envelope
is well resolved. At both blueshifted and redshifted velocities the emission appears to be circularly
symmetric, as is expected for a spherically expanding envelope. More interestingly, at velocities
between 59 -- 89 \kms, centered around the systemic velocity, the envelope is clearly elongated in
approximately the North-East to South-West direction. By fitting a two-dimensional Gaussian
to the intensity distribution of the $^{12}$CO J=2$-$1 emission at systemic velocity
V$_{\rm LSR}$ = 74 \kms, we obtain a position angle for the $^{12}$CO J=2$-$1 envelope of PA=70$^\circ$. 
The orientation of the
CO emission is consistent with the elongation found by Castro-Carrizo et al. (2007). The position-velocity (PV)
diagrams of the $^{12}$CO J=2--1 emission along the major (PA $\sim$70$^\circ$) and minor (PA $\sim$160$^\circ$) axis of the
envelope as presented in Figure 7 are typical of an expanding envelope at an expansion velocity of $\sim$40 \kms.

We show in Figure. 4 the channel maps of the $^{12}$CO J=2$-$1 emission, which is obtained by combining the
data from both compact and extended configurations of the SMA. The $^{12}$CO J=2$-$1 emission is
further resolved into more complex structures. In the velocity channels between 59 -- 89 \kms, centered
around the systemic velocity,
the $^{12}$CO J=2--1 emission consists of a central prominent hollow shell structure of $\sim$1\arcsec\, to 2\arcsec\,
in radius and a more clumpy arc or shell located between radii of 3\arcsec\, to 6\arcsec, which is more prominent in the South West quadrant of
the envelope. Close inspection of the velocity channels around
the systemic velocity V$_{\rm LSR}$ $\sim$ 74 \kms indicates that the central hollow shell-like structure is clumpy and shows
stronger emission toward the South-Western side.
The outer arc or shell of emission can be seen more easily in the azimuthal average of the $^{12}$CO J=2$-$1 brightness
temperature as shown in Figure. 13. We can clearly see a central depression within a radius of $\sim$ 1" and enhanced emission between
the radii of 3\arcsec\, to 6\arcsec. 
\subsection{$^{13}$CO J=$2-$1 emission}

The channel maps of $^{13}$CO J=2--1 emission obtained from the compact configuration data at
an angular resolution of $\sim$3\arcsec.5
is shown in Figure 5. The emission in the $^{13}$CO J=2--1 line appears fainter and more compact then seen in the
main isotope $^{12}$CO J=2--1 line. 
Like in $^{12}$CO J=2--1 emission, near the systemic velocity, the envelope traced by
$^{13}$CO J=2--1 emission is elongated at position angle of PA $\sim$
70$^\circ$. The centroid of the emission appears to be slightly offset from the stellar position. 
Also like in $^{12}$CO J=2--1 emission, at higher velocities
in both blue-shifted and red-sfhited parts the $^{13}$CO J=2--1 emission again 
shows a roughly circularly symmetric morphology,
as would be expected for a spherically expanding envelope. The spatial kinematics of the $^{13}$CO J=2--1 emission
can be more clearly seen in the position-velocity diagram (Figure 7) along the major axis (PA $\sim$70$^\circ$) of
the envelope. In the PV diagram, there is a small velocity gradient in the velocity range between 70 to 100 \kms. 
Closer inspection of the channel maps suggests that the small velocity gradient corresponds to the slight shift 
of the emission centroid toward the North-East quadrant of the envelope in the abovementioned velocity range.

\subsection{SO J$_{\rm K}$=6$_5$--5$_4$ emission}

In Figure. 6 we show channel maps of the SO 6$_5-$5$_4$ emission obtained from the data of the compact configuration. 
The emission is well resolved
in some of the velocity channels. A close inspection reveals that the distribution of SO 6$_5-$5$_4$ emission is different from that
of $^{12}$CO J=2$-$1. At extreme redshifted velocities the centroid of the emission is shifted to the East, whereas
at extreme blueshifted velocities the centroid of the emission is shifted to the West. 
In addition, the SO 6$_5-$5$_4$ emission in velocity channels centered around the systemic velocity is
clearly elongated to the South-West. This elongation is very similar to that seen in $^{12}$CO J=2$-$1 emission
at the same angular resolution (see Figure 3). More interestingly, at the opposing redshifted velocities of
84 to 102 \kms (see Figure 6) the emission is very compact and shifts to the North-East of the central
star. Such positional shift is shown more clearly in the position-velocity diagram of the
SO J$_K$=6$_5$--5$_4$ emission along the major axis (PA $\sim$70$^\circ$)
of the envelope. There is a clearly velocity gradient in the velocity range 
between 70 to 90 \kms, which is similar but more pronounced
that that seen in $^{13}$CO J=2--1 line. 

Thus, the SO 6$_5-$5$_4$ emission in particular but also the $^{13}$CO J=2$-$1
emission seem to better trace the asymmetric structure inside the envelope.
The small velocity gradient seen in both lines points to the presence of an ejecta in the envelope of
IRC+10420. From the abovementioned elongation of the envelope and the orientation of the positional velocity shift, 
we estimate that the
ejecta is oriented at position angle of PA=70$^\circ$.
We note that SO emission is known to be enhanced in AGB and post-AGB envelopes where collimated high velocity
outflows are present. 
Examples can be found in the rotten egg nebula OH 231.8+4.2 (S\'{a}nchez Contreras et al. 2000)
and VY CMa (Muller et al. 2007). In both cases SO emission is found to be strongly enhanced in the bipolar lobes. 
The SO enhancement in bipolar outflows could be related to the
synthesis of SO in the shocked molecular gas in these outflows. Alternatively, the warm environment around
supergiants like VY CMa and IRC+10420 could facilitate the synthesis of SO through chemical pathways such as 
S + OH $\longrightarrow$ SO + H or HS + O $\longrightarrow$ SO + H (Willacy \& Millar 1997).
\section{Structure of the envelope}
In this section we attempt to build a model for the circumstellar
envelope of IRC+10420 using our newly obtained SMA data and also the 
large body of observational data on IRC+10420, ranging from optical to milimeter wavelengths.
Our goal is to better understand the physical conditions of the molecular gas inside the
envelope and also to retrace the mass loss history of the central star.
The structure of the envelope around IRC+10420 has been previously modelled in the work of Castro-Carrizo et al. (2007),
by prescribing a radial distribution for the temperature in order to derive 
the gas density and consequently the mass loss rate. 
In our model we take into account the properties of
the dust inside envelope and the balance of heating and cooling processes. From the discussion in the previous
section, the envelope can be approximated as spherically symmetric.
The physical conditions such as kinetic temperature, density in the envelope can then be inferred from
detailed calculations of the energy balance within the envelope and matching the predictions of the radiative
transfer model to the observed strength of the CO rotational lines. This procedure represents an improvement in
comparison to previous models, and allows a better understanding of the physical conditions 
in the envelope of IRC+10420.

\subsection{A simple model for dust continuum emission}

To determine the heating process in the detected molecular shells requires the knowledge of the
momentum transfer coefficient (or the flux average extinction coefficient $<Q>$) between the dust 
particles and the gas molecules.
To estimate this quantity, we follow the treatment of Oudmaijer et al. (1996) who modelled in detail
the spectral energy distribution (SED) of IRC+10420. We emphasize here that by following
the treatment of Oudmaijer et al. (1996), the structure of the molecular envelope as seen in our SMA data and
previously in the data of Castro-Carrizo et al. (2007) is not considered explicitly.
 
We use the photometric data of IRC+10420 collected
by Jones et al. (1993) and Oudmaijer et al. (1996), i.e. the 1992 photometric dataset. 
We also assume that the dust consists of 
silicate particles with a radius of $a$ = 0.05 $\mu$m. The dust
opacity as a function of wavelength is taken from the work of Volk \& Kwok (1988). Following Oudmaijer et al. (1996),
we scale the opacity law to match the absolute value of the opacity at 60$\mu$m of
$\kappa_{60 \mu m}$ = 150 cm$^2$ g$^{-1}$. We also a use a specific density for dust particles of
$\rho$ = 2 g cm$^{-3}$ and assume a dust to gas ratio  
$\Psi$ = 5 10$^{-3}$ as used by Oudmaijer et al. (1996).

To fit the SED of IRC+10420, especially in the mid-IR region between 5 to 20 $\mu$m, Oudmaijer et al. (1996)
suggest that two dust shells are needed: a hot inner shell with a low mass loss rate and an outer cooler shell 
with a higher mass loss rate. We note that Bl\"{o}cker et al. (1999) 
also reached same conclusion in their attempt to fit the SED of IRC+10420.
In our model we also use two dust shells: a hot inner shell with a low mass loss rate of \.{M} = 8 10$^{-5}$ M$_\odot$/yr 
located between 2.6 10$^{15}$ cm (70 stellar radii) to 1.8 10$^{16}$ cm (486 stellar radii) and
an outer cooler shell with a higher mass loss rate of \.{M} = 1.2 10$^{-3}$ M$_\odot$/yr extending to a radial
distance of 10$^{18}$ cm where the dust particle number density drops to very low values.
We also infer a stellar luminosity of 6 10$^5$ L$_\odot$. The size of the dust shells, the mass loss 
rates and the stellar luminosity are very similar to that derived by Oudmaijer et al. (1996) scaled to the adopted distance of 5 kpc. 
The results of our radiative transfer model are presented in Figure 8 and 9. The fluxes from the model are corrected
for an interstellar extinction of A$_{\rm V}$=5 as suggested by Oudmaijer et al. (1996). As can be seen in 
Figure 8, the results of our model are in close agreement with that shown in Oudmaijer et al. (1996) and match 
reasonably the observed SED of IRC+10420. As shown in Figure. 9, the dust temperature
decreases monotonically from $\sim$1000 K at the inner radius to $\sim$50 K at the outer radius
of the dust envelope, even though there is a jump in mass loss rate between the inner hot dust shell and the
outer cooler dust shell at a radial distance of 1.8 10$^{16}$ cm. 

From the flux $F_\lambda$ of the radiation field at each point in the envelope we can define the flux average 
extinction coefficient $<Q>$, which characterizes the transfer of angular momentum from radiation photons to dust
particles during the absorption and reemission processes of the continuum radiation. Because in our case 
we do not consider the scattering of radiation, the extinction coefficient $Q_\lambda$ is 
simply the opacity $\kappa_\lambda$, where $\lambda$ is the wavelength of the radiation.
\begin{equation}
<Q> = \frac{\int_{0}^{\infty} F_\lambda Q_\lambda d\lambda}{\int_{0}^{\infty}F_\lambda d\lambda} = 
\frac{\int_{0}^{\infty} F_\lambda \kappa_\lambda d\lambda}{\int_{0}^{\infty}F_\lambda d\lambda}
\end{equation}
The flux average extinction coefficient $<Q>$ 
changes significantly in the inner part of the dust envelope but 
reaches a constant value in the outer part. Because the molecular shells are located in the outer 
part of the dust envelope, at radii larger than a few times 10$^{16}$ cm, 
for simplicity, we will adopt a constant value $<Q>$=0.025 (see Figure 9).

\subsection{Excitation of CO molecule}
The temperature of the molecular gas at any point in the envelope is determined by the balance between cooling
due to adiabatic expansion and molecular emissions, and the heating, which is mainly due to collisions of molecules
with dust grains (Goldreich \& Scoville 1976). As the dust grains stream through the 
gas under radiation pressure, the grains transfer angular momentum to
the gas by collisions. The heating term is therefore
directly related to the drift velocity:
\begin{equation}
H  =  \frac{1}{2}\rho \pi a^2 n_d v_{\rm drift}^3
\end{equation} 
where $n_d$ is the number density of dust particles. The drift velocity is determined by
the relation:
\begin{equation}
v_{\rm drift} = \left(\frac{L_{*}V_{\rm exp}<Q>}{\dot{M}c}\right)^{1/2}
\end{equation}
where $L_*$ is the stellar luminosity, $V_{exp}$ is the expansion velocity, $\dot{M}$ is the mass loss rate,
and $c$ is the speed of light, $<Q>$ is the flux average extinction coefficient of dust particles estimated
in previous section. The same dust to gas ratio of $\Psi$ = 5 10$^{-3}$ is used to calculate the heating rate.

Generally, in oxygen-rich envelopes the main coolants are H$_2$O and CO molecules.  
The contribution of H$_2$O molecules to the cooling process in IRC+10420 is quite uncertain.
We note that far-IR emission lines of H$_2$O are not detected in the ISO data (Molster et al. 2002). 
In addition, OH masers are seen in a shell of radius 1 to 1.5 arcsec (Nedoluha \& Bowers 1992). 
Because OH radical is the photodissociation product
of H$_2$O, it is likely that H$_2$O molecules can only exist at even smaller radii, i.e. inside the central
cavity disccused previously.  Therefore, we do not expect the H$_2$O molecules to be
present in significant quantity in the part of the molecular envelope traced in CO emission
As a result, we do not include the cooling due to the H$_2$O molecule in our model
and calculate the cooling solely due to CO molecule and its isotopomer $^{13}$CO. 

We use the thermal balance and radiative transfer model published by Dinh-V-Trung \& Nguyen-Q-Rieu (2000) to
derive the temperature profile of the molecular gas and to
calculate the excitation of CO molecules and predict the strength of CO rotational transitions. We take
into account all rotational levels of CO molecules up to J=20. The collisional cross sections with H$_2$
are taken from Flower (2001), assuming an ortho to para ratio of 3 for molecular hydrogen. The local 
linewidth is determined by the turbulence velocity and the local thermal linewidth. Because the expansion
velocity of the gas in the envelope of IRC+10420 is large and the observed CO line profiles have smooth
slopes at both blueshifted and redshifted edges (see Figure 11), we need to assume a constant
turbulence velocity of 3 \kms. We note that this is higher than normally used (about 1 \kms) for the envelope around AGB stars where
the expansion velocity is lower (Justtanont et al. 1994). In our model, the spherical envelope is covered with a 
grid of 90 radial points. For each rotational 
transition, the radiative transfer equation is integrated accurately through the envelope
to determine the average radiation field at each radial mesh point. We use the approximate lambda operator
method together with Ng-acceleration to update the populations of CO molecules of different energy levels. 
Once convergence is achieved, we convolve the predicted radial distribution of the line intensity with 
a Gaussian of specified FWHM to produce simulated data for comparison with observations.

As discussed in the previous section, there is a clear evidence for the presence of two molecular shells from
our SMA data. The size and location of each molecular shell are determined from fits to the interferometric data.
In our model we adjust both the mass loss rates and the relative abundance of CO molecules in the two molecular shells 
to match both the single dish and interferometric data. 
We find that a relatively low mass loss rate similar to that
derived by Castro-Carrizo et al. (2007) would result in a too high gas temperature in the envelope as the heating term
becomes more dominant in comparison to the cooling term.
We also find that a high
abundance of CO of a few times 10$^{-4}$ as usually used for oxygen rich envelopes 
(Kemper et al. 2003, Castro-Carrizo et al. 2007) 
would result in CO lines that are too strong. Instead, a relative abundance of 10$^{-4}$ for CO molecules with
respect to H$_2$ for the CO molecules provides reasonable fit to all the available data.
The radial profile of the gas temperature is shown in 
Figure 10. The sudden jump in gas temperature at the
inner radius of the second shell ($\sim$1.85 10$^{17}$ cm) is due to the change in the mass loss rate. We can see that
the gas temperature in the envelope of IRC+10420, although lower than assumed in Castro-Carrizo et al. (2007), is
higher than that typically found in circumstellar envelopes. Even at the outermost radius of the envelope, the
gas temperature is still around 50 K. For comparison, in the envelope around OH/IR stars, 
the gas temperature at similar radial distances from the central star is predicted to be much lower, 
10 K or even less (Justtanont et al. 1994,
Groenewegen 1994). The main reason for the elevated gas temperature in the envelope of IRC+10420 is the
strong heating due to the enormous luminosity of the central yellow hypergiant. The cooling due to molecular
emissions is also affected by the lower abundance of CO, that we find neccessary to match the data.
A summary of the parameters used in our model is presented in Table 3.

As shown in Figures 11--14, the results of our model are in reasonable agreement with both the strengths of CO lines measured by
single dish telescopes and the radial distribution of brightness temperature of J=1--0 and J=2--1 lines
observed with interferometers. We note that except for the extra emission in the blue part of the
line profile between 65 and 75 \kms, the shape of the J=2--1 and J=3--2 is in close agreement with observation.
Higher transitions J=4--3 and 6--5 show some evidence of narrower linewidth than predicted by the model.
One possible explanation is that the expansion velocity in the inner region traced by these high lying transitions has
slightly lower expansion velocity than in the outer part of the envelope.

By comparing the model predictions with the strengths of $^{13}$CO J=1--0 and J=2--1 lines observed with
the IRAM 30m telescope (Bujarrabal et al. 2001), we derive a relative abundance for $^{13}$CO/H$_2$ of
1.5 10$^{-5}$, or an isotopic ratio $^{12}$C/$^{13}$C $\sim$ 6. This ratio is very similar to
the values $^{12}$C/$^{13}$C = 6 and 12, respectively, found in the red supergiants $\alpha$ Ori and $\alpha$ Sco
(Harris \& Lambert 1984, Hinkle et al. 1976). Such a low isotopic ratio suggests
that IRC+10420 has experienced significant mixing of H burning products to its surface prior to
the ejection of the material in the envelope.

The difference between the mass-loss rates and gas temperature profile derived from our modelling and that
obtained by Castro-Carrizo et al. (2007) can be understood as the consequence of the higher gas temperature 
and higher CO abundance adopted in their work. For the same strength of the CO rotational lines, the higher CO abundance 
can be almost exactly compensated by a corresponding decrease in the mass loss rate.

Our results indicate the presence of two separate shells (shell I and shell II, see Table 3) with slightly 
different mass loss rates. With an expansion velocity of 38 \kms, the time interval 
between the two shells is $\sim$ 200 yrs. The cavity inside shell I
also implies a dramatic decrease in mass loss from IRC+10420 over the last $\sim$300 yrs. Thus IRC+10420
loses mass in intense bursts, separated by relatively quiet periods of a few hundreds years in duration.
\begin{table}[h] \begin{center} \begin{tabular}{lll}
\multicolumn{3}{l}{Table 3: Parameters of the model for IRC+10420 envelope.} \\
\hline
\hline
Parameters                         & Shell I     & Shell II       \\
\hline
Inner radius  & 3.5 10$^{16}$ cm & 1.85 10$^{17}$ cm \\
Outer radius  & 1.5 10$^{17}$ cm & 5 10$^{17}$ cm \\
Mass loss rate \.{M} & 9 10$^{-4}$ M$_\odot$ yr$^{-1}$ & 7 10$^{-4}$ M$_\odot$ yr$^{-1}$ \\
Expansion velocity V$_{\rm exp}$ & 38 kms$^{-1}$ & 38 kms$^{-1}$ \\
Dust to gas ratio $\Psi$ & 5 10$^{-3}$ & 5 10$^{-3}$ \\
Momentum transfer coefficient $<Q>$ & 0.025 & 0.025 \\
$^{12}$CO/H$_2$ & 10$^{-4}$ & 10$^{-4}$ \\
$^{13}$CO/H$_2$ & 1.5 10$^{-5}$ & 1.5 10$^{-5}$ \\
Turbulent velocity & 3 kms$^{-1}$ & 3 kms$^{-1}$ \\
\hline
\end{tabular} \end{center} \end{table}

\subsection{Caveats}
In our model we consider separately the dust and gas component in the envelope of IRC+10420. The dust continuum
emission model is used as a simple way to estimate the $<Q>$ parameter of dust particles, which is needed for the thermal balance 
calculation in the molecular shells. It turned out that the mass loss rates of the molecular shells between
9 10$^{-4}$ M$_\odot$/yr for shell I and 7 10$^{-4}$ M$_\odot$/yr for shell II are lower but quite close to the 
mass loss rate of 1.2 10$^{-3}$ M$_\odot$/yr of the cool outer dust shell inferred from the modelling 
the SED of the envelope. Our approach, although less ad hoc than simply assuming a value of the $<Q>$ parameter for dust particles,
is not fully self-consistent in the treatment of the dust and gas component.
It would be desirable in the future to treat both the dust and gas component together when high angular resolution
data on the dust continuum emission and higher J lines of CO become available.

Another caveat of our model is the assumption that the heating of molecular gas is mainly due to collision between dust
particles and gas molecules. The high luminosity of the central star of IRC+10420 suggests that 
radiation pressure on dust particles is a possible mechanism for driving the
wind as assumed in our mode. However, given the high expansion
velocity and the presence of many small scale features in high angular resolution optical images of the envelope
(Humphreys et al. 1997), which has been interpreted as jets or condensations, other mechanisms
such as shocks produced by interaction between higher velocity ejecta and the envelope should be considered.
The heating due shocks within the envelope might contribute to the thermal balance of the molecular gas in IRC+10420.

\section{Comparison with optical imaging data}
The circumstellar envelope around IRC+10420 is known to be very complex, with a number of
peculiar structures (Humphreys et al. 1997). In the outer part of the envelope at radial distance
of $\sim$3\arcsec\, to 6\arcsec, several arc-like features can be seen. These features have been interpreted as
representing different mass loss episodes of IRC+10420. In addition, condensations or blobs are seen
closer (about 1\arcsec\, to 2\arcsec\, in radius) to the central star. That might represent a more recent mass loss
episode. These features correspond spatially to the two molecular shells (shell I and II) identified in our 
observations. In addition, Humphreys et al. (1997) also identified several broad fan-shaped features to the
South-West side of the central star, between radius of 0".5 to 2". From their higher surface brightness in comparison
with other parts of the envelope,
Humphrey et al. (1997) suggest that these features are ejecta moving obliquely toward the observer.
As discussed in previous section, the channel maps of $^{13}$CO J=2--1 and SO J$_{\rm K}$=6$_5$--5$_4$ (Figures 5 and 6)
reveal enhanced emissions in blueshifted velocity channels together with a positional shift of the emission centroid 
also to the South-West quadrant of the envelope. Therefore, the spatial kinematics obtained from our observations and 
the properties of these fan-shaped features are consistent with the presence of an ejecta in the South-West
side of the envelope around IRC+10420. 

\section{Summary}

We have used the sub-millimeter array to image and study the structure of the molecular envelope
around IRC+10420. Our observations reveal a large expanding envelope with a clumpy and very complex
structure. The envelope shows clear asymmetry in $^{12}$CO J=2--1 emission in the South-West direction
at position angle PA$\sim$70$^\circ$. The elongation of the envelope is found even more pronounced in
the emission of $^{13}$CO J=2--1 and SO J$_{\rm K}$=6$_5$--5$_4$. A positional shift with velocity is
seen in the above emission lines, suggesting the presence of a weak bipolar outflow in the envelope of
IRC+10420. 

In the higher resolution data of $^{12}$CO J=2--1, we find that 
the envelope has two components: (1) a inner shell (shell I) located between radius
of about 1"-2"; (2) an outer shell (shell II) between radius 3" to 6". These shells represernt two previous 
mass loss episodes from IRC+10420. 

We constrain the physical conditions inside the envelope
by modelling the dust properties, the heating and cooling of molecular gas. From comparison with observations
we derive the size and the location of each molecular shell. We estimate
a mass loss rate of $\sim$ 9 10$^{-4}$ M$_\odot$ yr$^{-1}$ for shell I and 7 10$^{-4}$ M$_\odot$ yr$^{-1}$ for
shell II. The gas temperature is found to be an usually high in IRC+10420 in comparison with other oxygen rich envelopes,
mainly due to the large heating induced by the large luminosity of the central star.

We also derive a low isotopic ratio $^{12}$C/$^{13}$C = 6 for IRC+10420, which suggests a strong mixing of processed
material from stellar interior to the surface of the star.

\acknowledgements

We are grateful to the Sub-millimeter array staff for their help in carrying out the observations. 
The SMA is a collaborative project between the Smithsonian Astrophysical Observatory and 
Academia Sinica Institute of Astronomy and Astrophysics of Taiwan.  
We thank F. Kemper
and D. Teyssier for providing single dish CO spectra of IRC+10420. This research has made use of 
NASA's Astrophysics Data System Bibliographic Services
and the SIMBAD database, operated at CDS, Strasbourg, France.

\newpage

\begin{figure}[h] \begin{center}
\includegraphics[width=10cm]{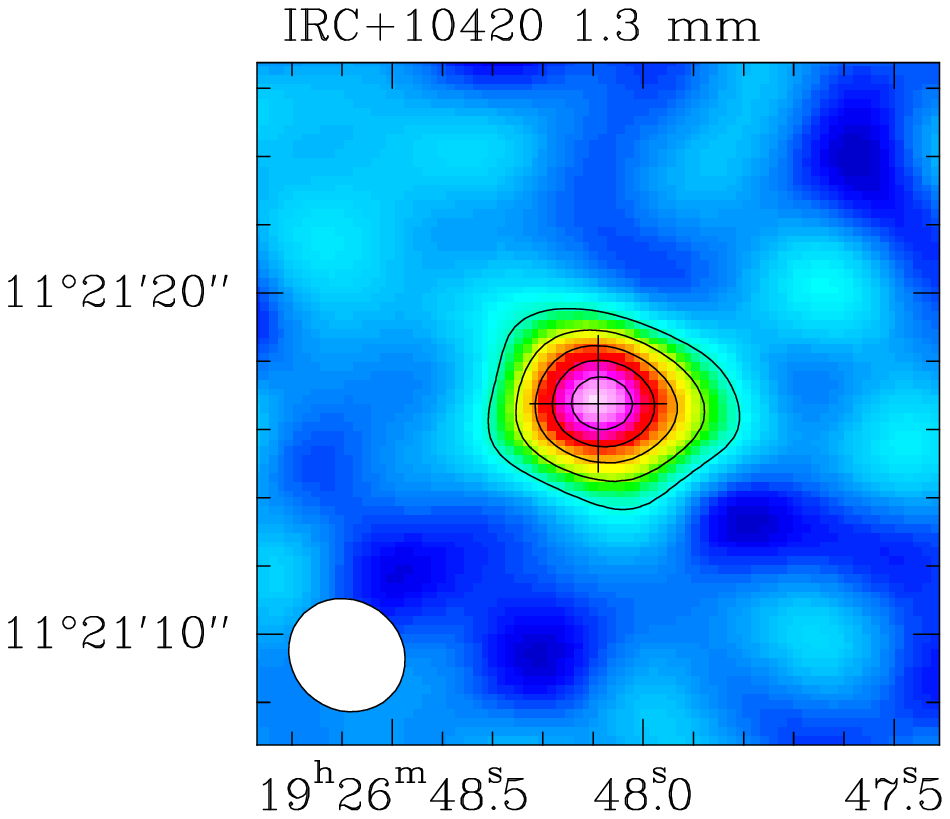}
\caption{1.3 mm continuum emission map of IRC+10420 obtained with the SMA compact configuration.
The synthesized beam is 3\arcsec.52 x 3\arcsec.21 (PA $= 54\degr$) as indicated by the ellipse
in the bottom left corner. Contour levels are every 6 mJy beam$^{-1}$ ($\sim 3\sigma$).}
\label{figcont}
\end{center} \end{figure}

\begin{figure}[h] \begin{center}
\includegraphics[height=14cm]{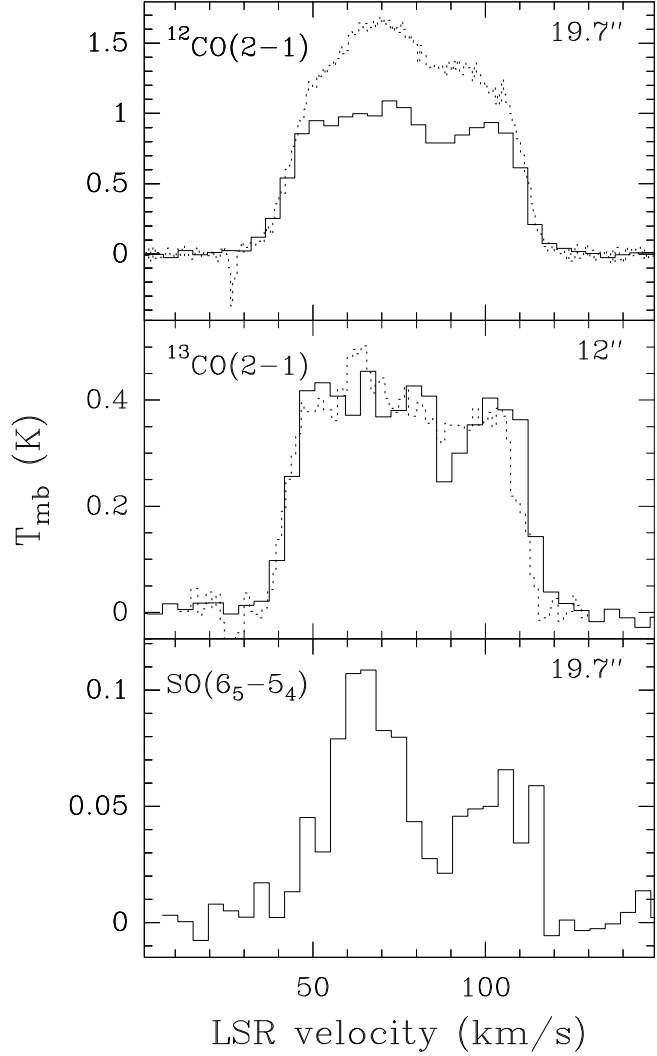}
\caption{Spectra of $^{12}$CO J=$2-$1 ({\em top}), $^{13}$CO J=$2-$1 ({\em middle}) and SO(6$_5-$5$_4$)
({\em bottom}) transitions towards IRC+10420. The spectra are shown in main beam brightness temperature
with a telescope beam as mentioned in top right corner of each box.
The dotted lines show the spectra of $^{12}$CO J=$2-$1 obtained with JCMT (Kemper et al. 2003)
and $^{13}$CO J=$2-$1 obtained with the IRAM 30m (Bujarrabal et al. 2001)}
\label{figspec}
\end{center} \end{figure}

\begin{figure}[h] \begin{center}
\includegraphics[height=11cm]{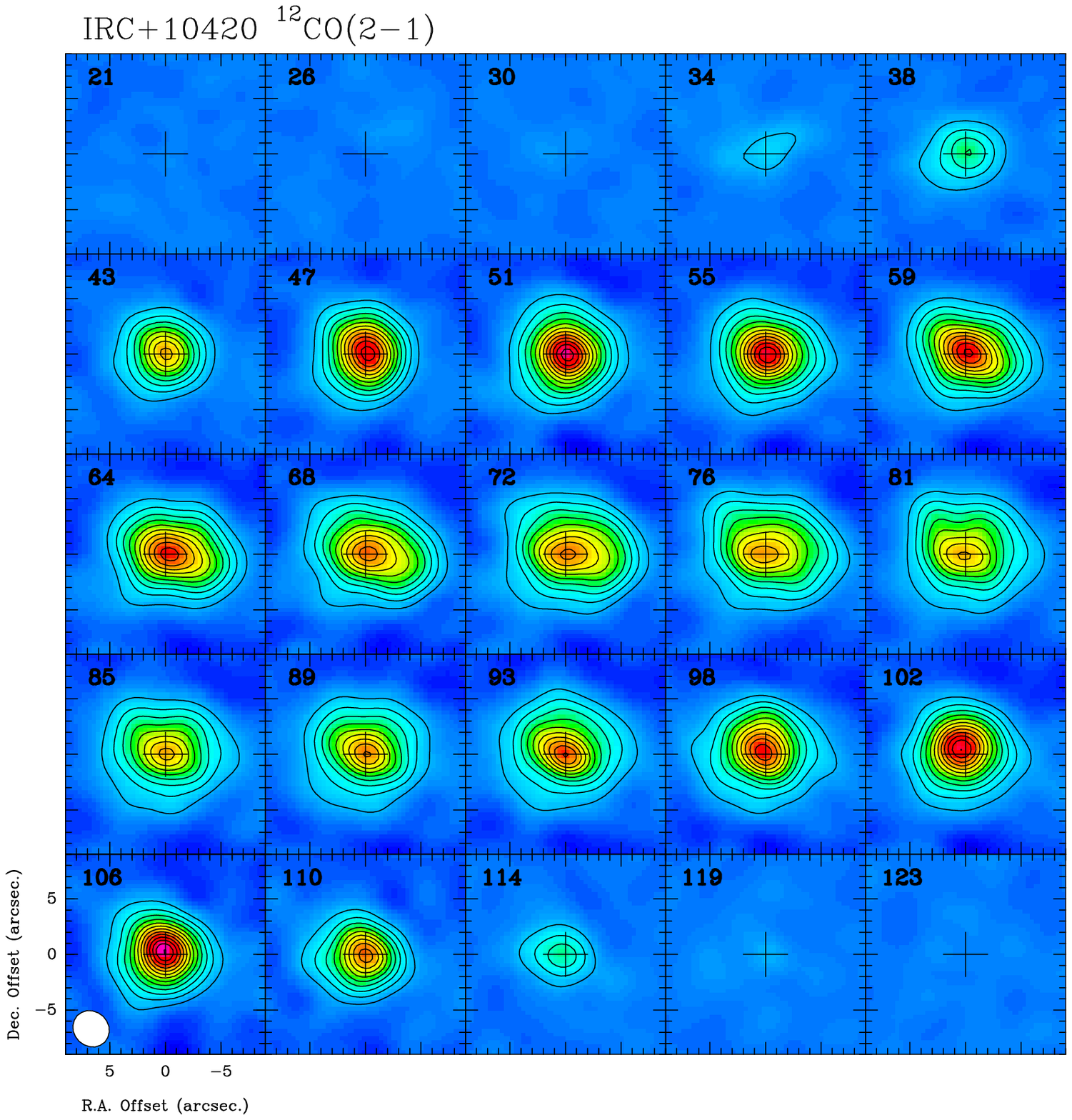}
\caption{Channel maps of the $^{12}$CO J=2--1 line emission obtained from the SMA compact configuration
alone. The synthesized beam is 3\arcsec.44 x 3\arcsec.12 (PA = 46$\degr$) and is indicated in the
bottom left corner of the figure. Contour levels are drawn every 0.6 Jy beam$^{-1}$ or  1.3 K ($\sim$ 10$\sigma$).
The velocity resolution is 4.2 km~s$^{-1}$. The cross indicates the position of the continuum peak emission.} 
\label{fig12co}
\end{center} \end{figure}

\begin{figure}[h] \begin{center}
\includegraphics[height=11cm]{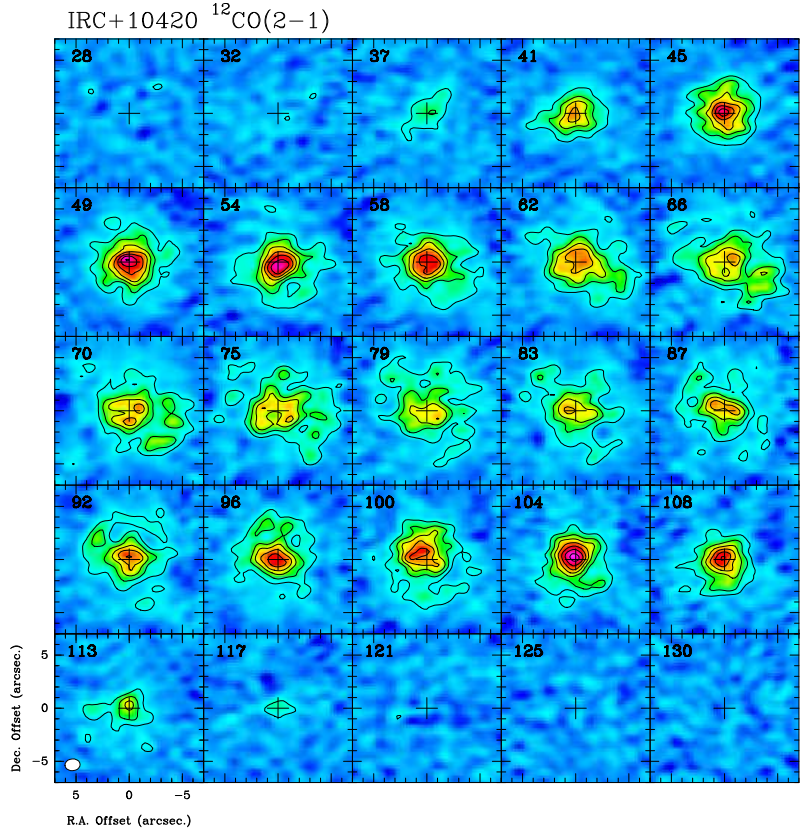}
\caption{Channel maps of the $^{12}$CO J=2--1 line emission obtained by combining data from compact and 
extended configurations. The synthesized beam is 1\arcsec.44 x 1\arcsec.12 (PA = 46$\degr$) and is
indicated in the bottom left corner of the figure. Contour levels are drawn every 0.24 Jy beam$^{-1}$ or 3.4 K
($\sim$ 3$\sigma$). The velocity resolution is 4.2 km~s$^{-1}$. The cross indicates the position of the
continuum peak emission.} 
\label{12cocomb}
\end{center} \end{figure}

\begin{figure}[h] \begin{center}
\includegraphics[height=11cm]{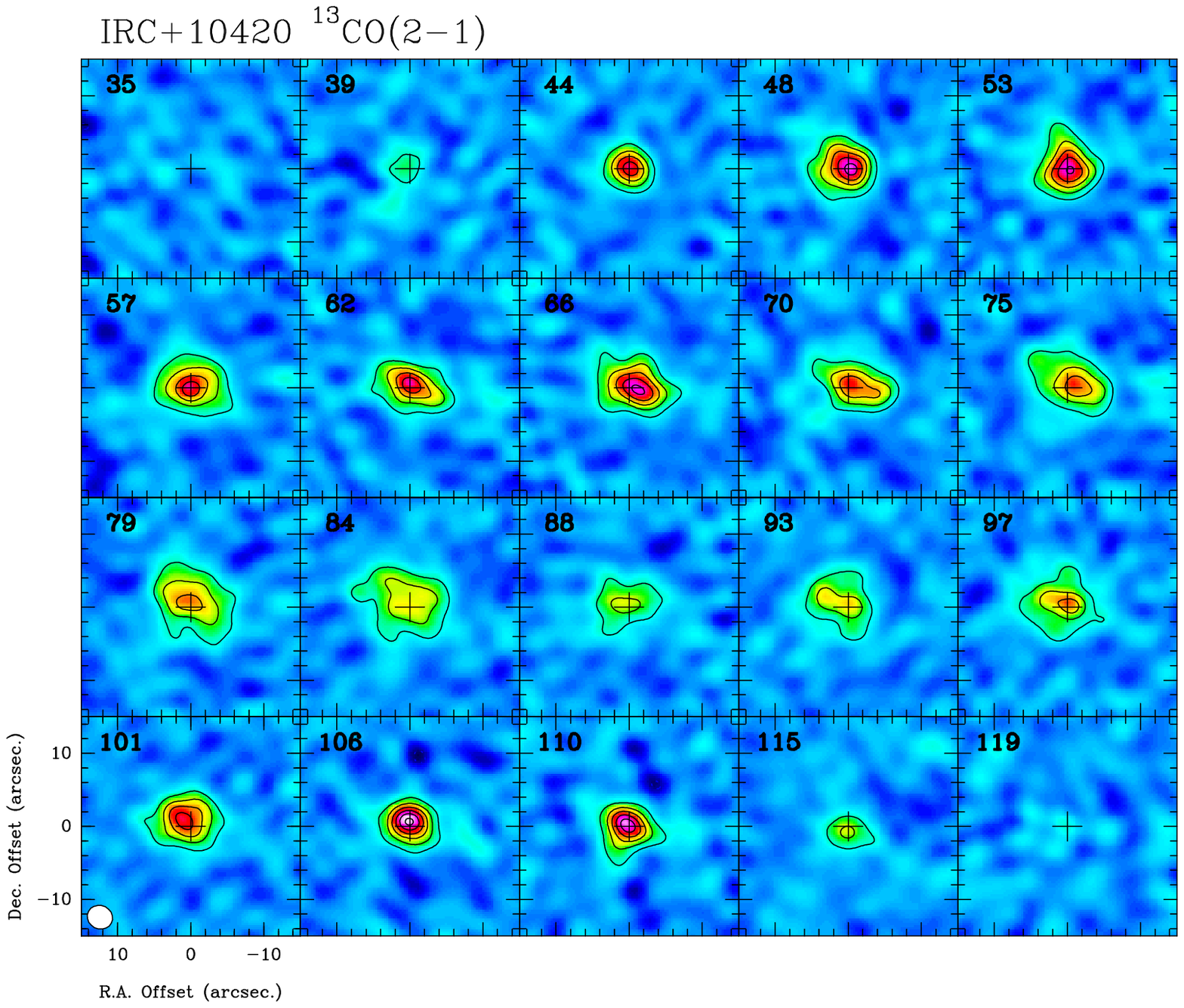}
\caption{Channel maps of the $^{13}$CO J=2--1 line emission obtained from the SMA compact configuration
alone. The synthesized beam is 3\arcsec.52 x 3\arcsec.21 (PA = 63$\degr$) and is indicated in the
bottom left corner of the figure. Contour levels are drawn every 0.2 Jy beam$^{-1}$ or 0.45 K ($\sim$ 4$\sigma$).
The velocity resolution is 4.4 km~s$^{-1}$. The cross indicates the position of the continuum peak emission.}
\label{fig13co}
\end{center} \end{figure}

\begin{figure}[h] \begin{center}
\includegraphics[height=11cm]{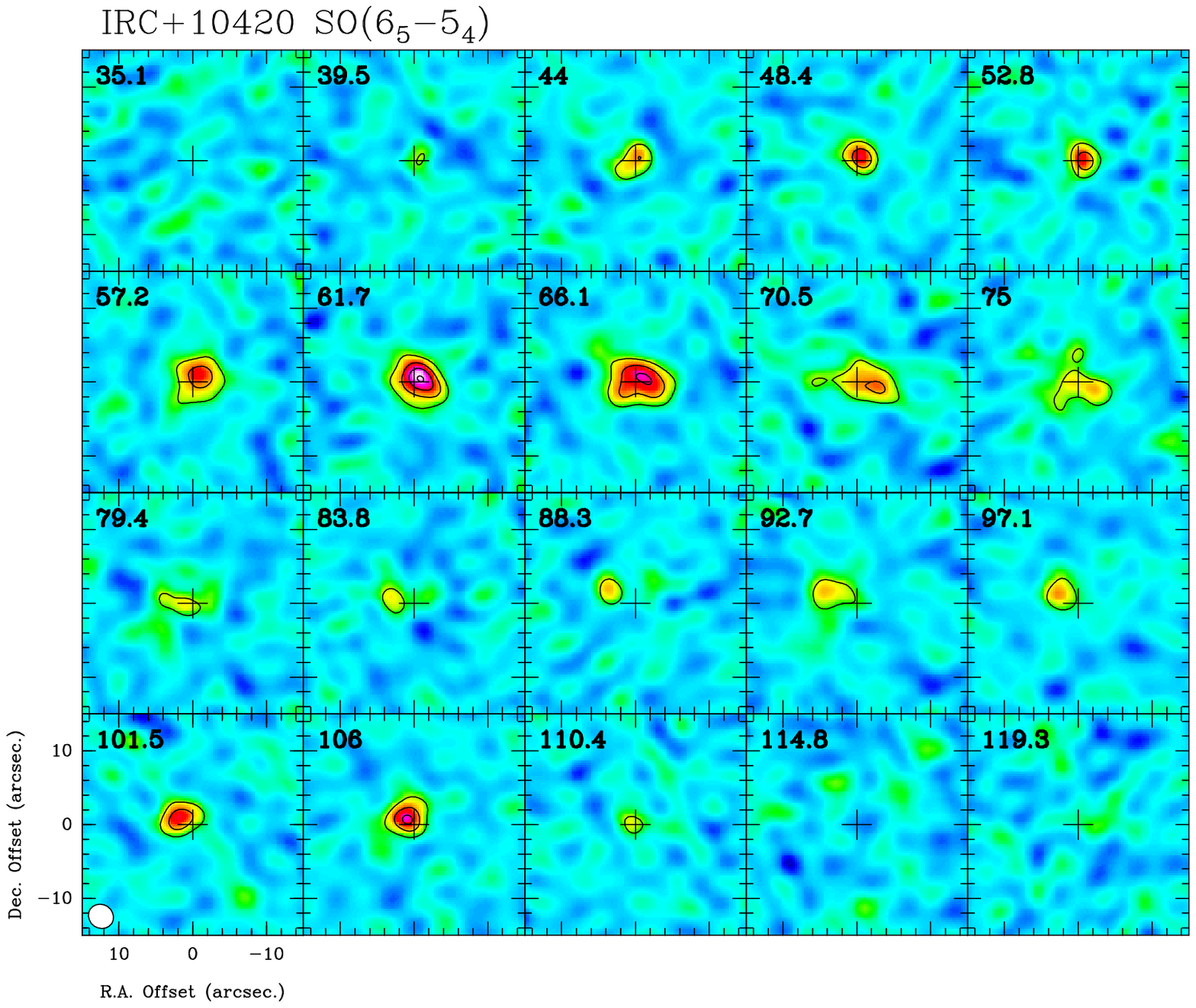}
\caption{Channel maps of the SO J$_{\rm K}$=6$_5$--5$_4$ line emission obtained from the SMA compact configuration
alone. The synthesized beam is 3\arcsec.51 x 3\arcsec.21 (PA = 63$\degr$) and is indicated in the
bottom left corner of the figure. Contour levels are drawn every 0.2 Jy beam$^{-1}$ or 0.45 K ($\sim$ 4$\sigma$).
The velocity resolution is 4.4 km~s$^{-1}$. The cross indicates the position of the continuum peak emission.}
\label{figso}
\end{center} \end{figure}

\begin{figure}[h] \begin{center}
\includegraphics[height=14cm]{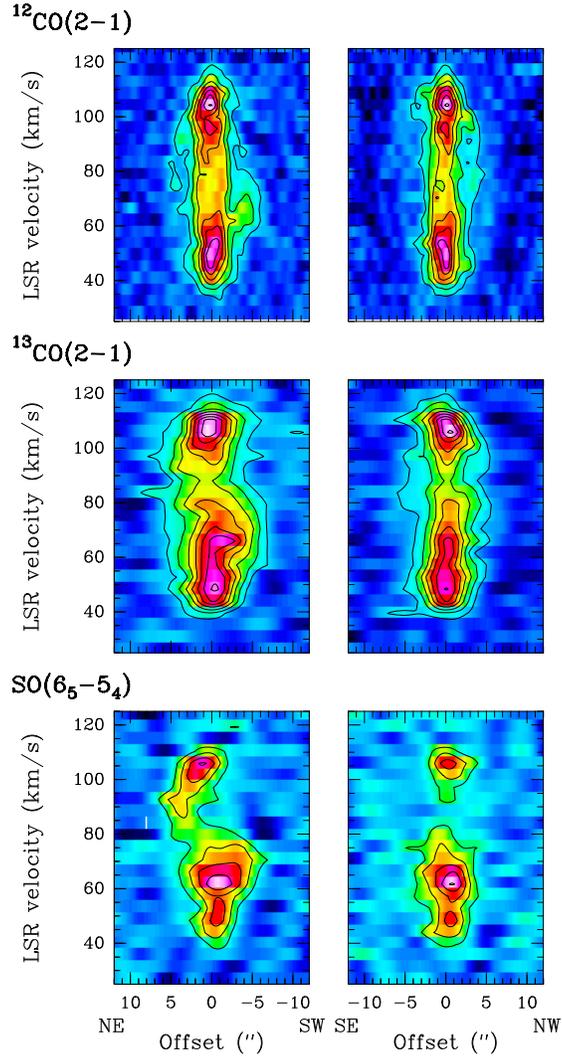}
\caption{Position-velocity diagrams along the major axis, defined at PA=70$^\circ$ ({\em left})
and perpendicularly along the minor axis ({\em right}). Contour levels are every 0.24 Jy~beam$^{-1}$
for $^{12}$CO J=2--1 and every 0.2 Jy~beam$^{-1}$ for $^{13}$CO J=2--1 and SO 6$_5$--5$_4$.} 
\label{pv}
\end{center} \end{figure}

\begin{figure}[h] \begin{center}
\setlength{\unitlength}{1cm}
\begin{picture}(15.,20.)
\put(0,12){\resizebox{16cm}{!}{\rotatebox{-90}{\includegraphics*{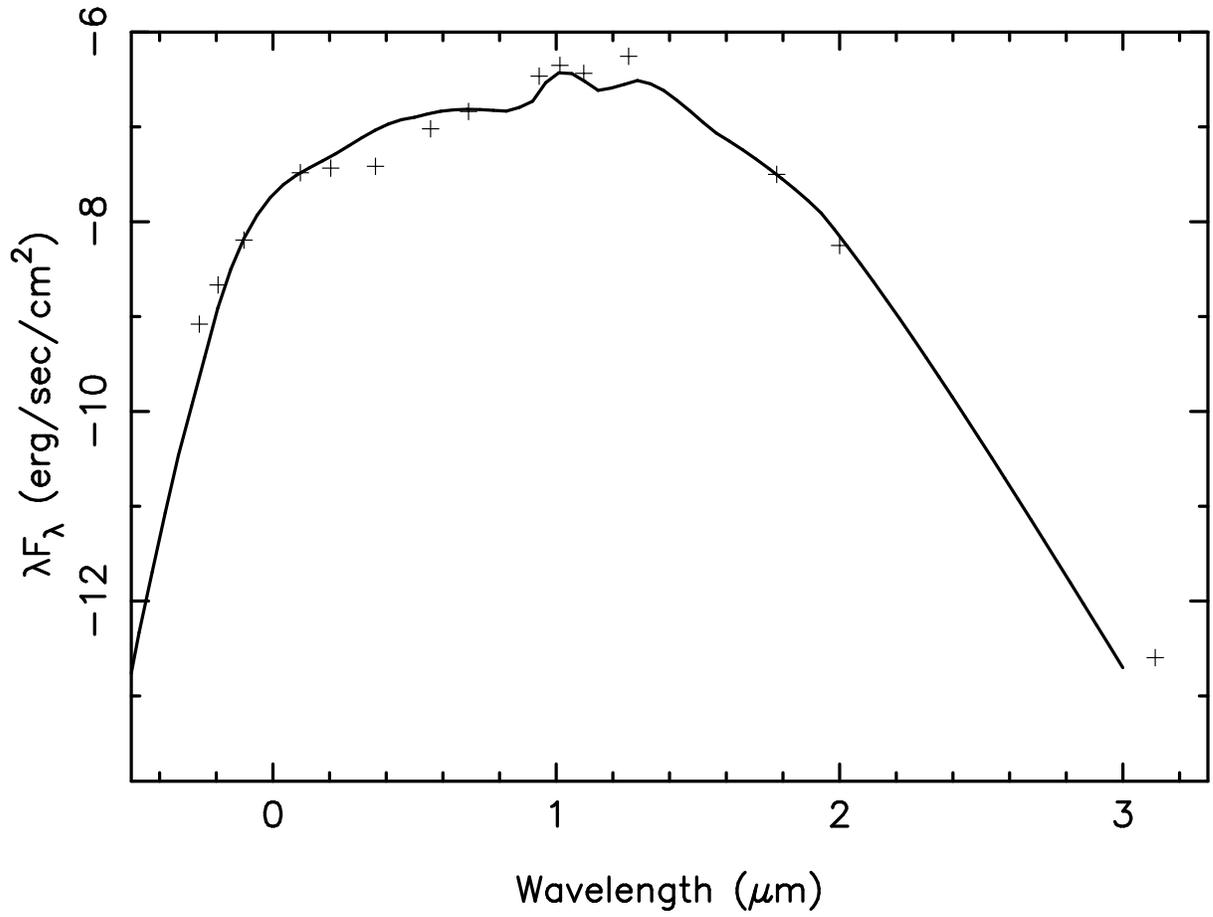}}}}
\end{picture}
\caption{Spectral energy distribution of IRC+10420. The crosses denote observational data from Jones et al. (1993)
and Oudmaijer et al. (1996).
The solid line represents the model SED.} 
\label{sed}
\end{center} \end{figure}

\begin{figure}[h] \begin{center}
\setlength{\unitlength}{1cm}
\begin{picture}(15.,20.)
\put(0,15){\resizebox{16cm}{!}{\rotatebox{-90}{\includegraphics*{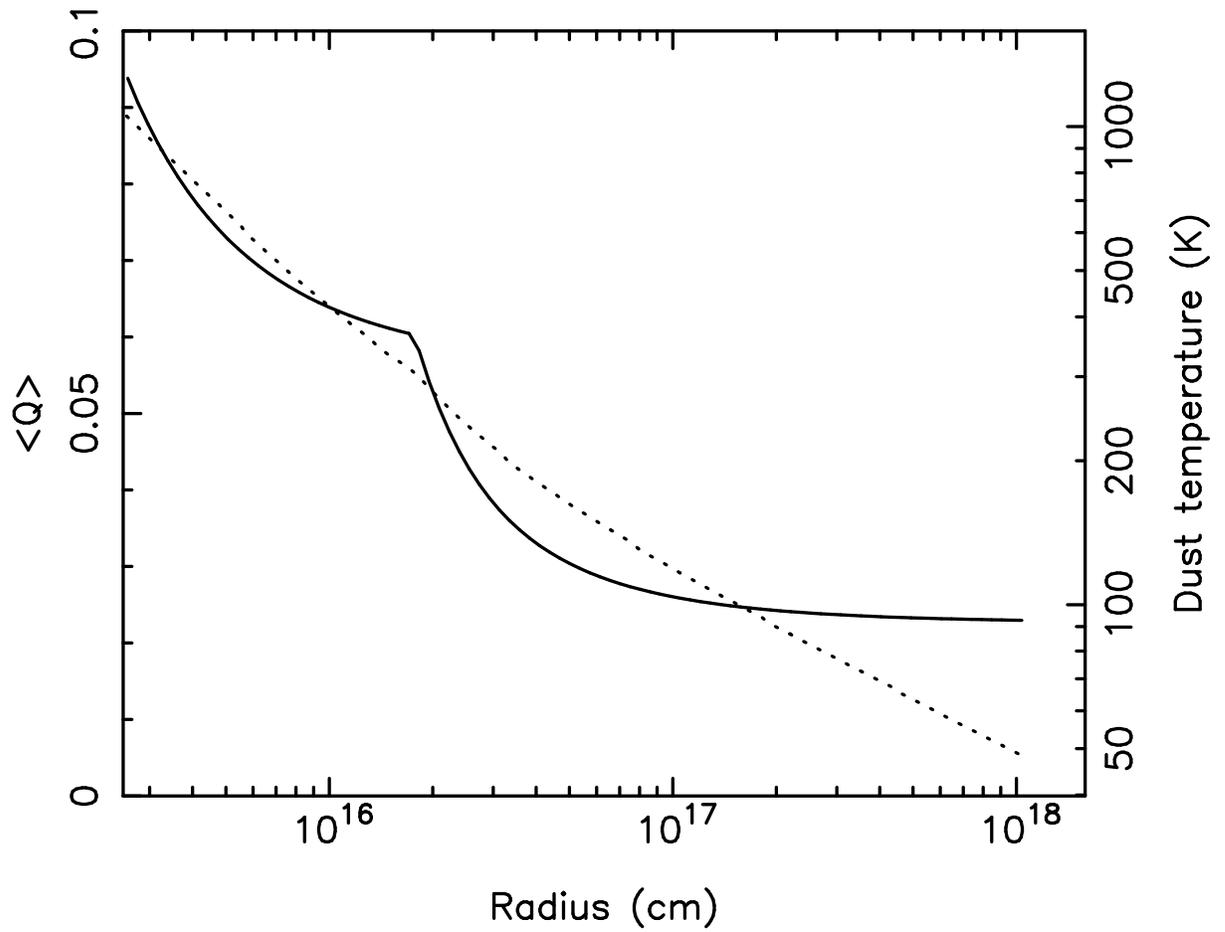}}}}
\end{picture}
\caption{Dust temperature profile (dotted line) and flux averaged extinction coefficient $<Q>$ (solid line)
in the envelope of IRC+10420 obtained from matching the model prediction of the SED to the observational
data.} 
\label{q-dtemp}
\end{center} \end{figure}

\begin{figure}[h] \begin{center}
\setlength{\unitlength}{1cm}
\begin{picture}(15.,20.)
\put(0,15){\resizebox{16cm}{!}{\rotatebox{-90}{\includegraphics*{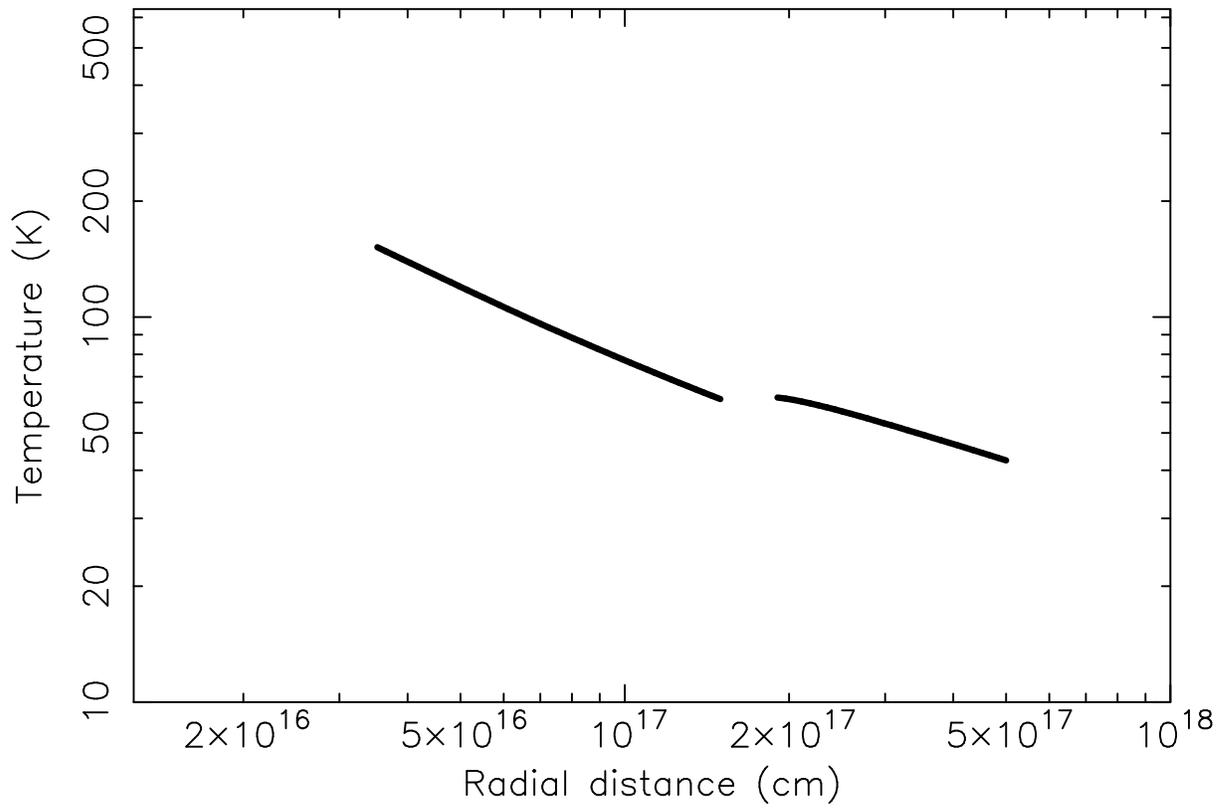}}}}
\end{picture}
\caption{The predicted radial profile of the molecular gas temperature in the envelope of IRC+10420.} 
\label{gas-temp}
\end{center} \end{figure}

\begin{figure}[h] \begin{center}
\includegraphics[height=18cm]{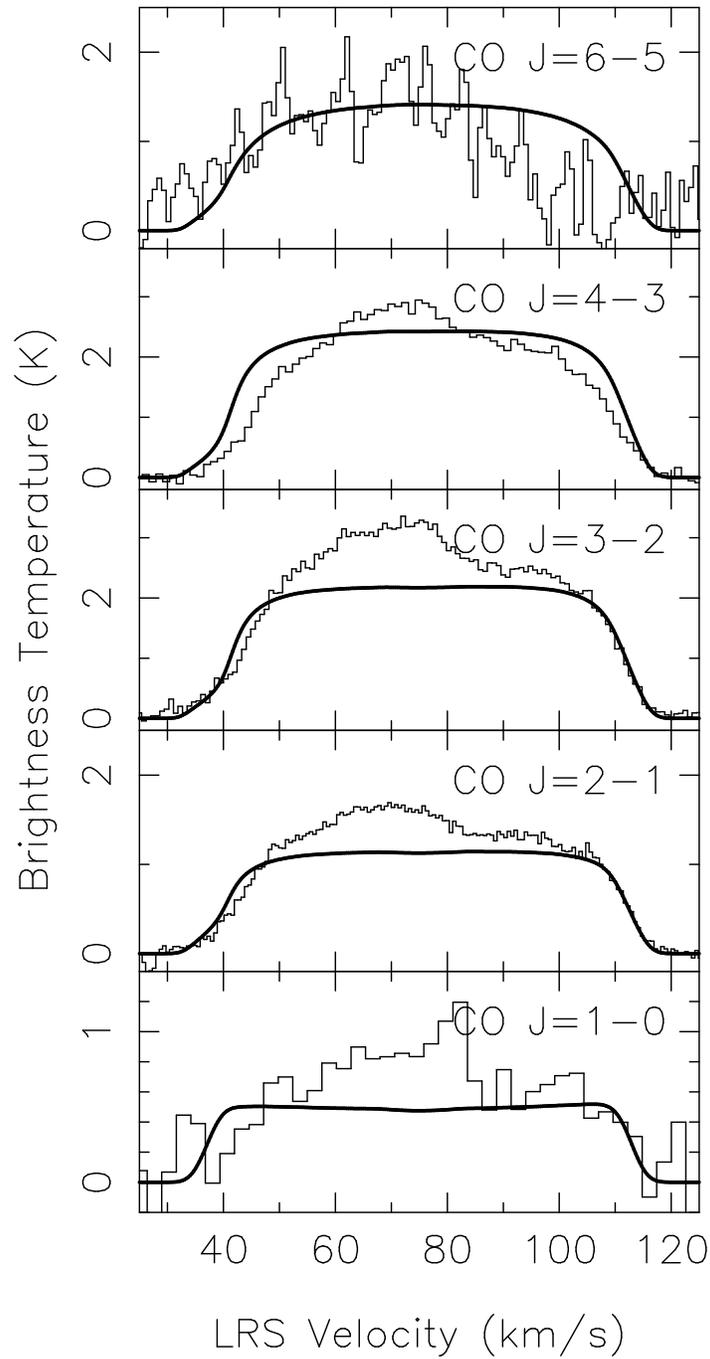}
\caption{Comparison between observations of CO transitions from Kemper et al. (2003) and Teyssier et al. (2006) 
and our model predictions. The
telescope beams are 22\arcsec\, for J=1--0, 19\arcsec.7 for J=2--1, 13\arcsec.2 for J=3--2, 10\arcsec.8 for J=4--3 and
10\arcsec.3 for J=6--5 line, respectively.} 
\label{copro}
\end{center} \end{figure}

\begin{figure}[h]
\setlength{\unitlength}{1cm}
\begin{picture}(15.,20.)
\put(0,12){\resizebox{15cm}{!}{\rotatebox{-90}{\includegraphics*{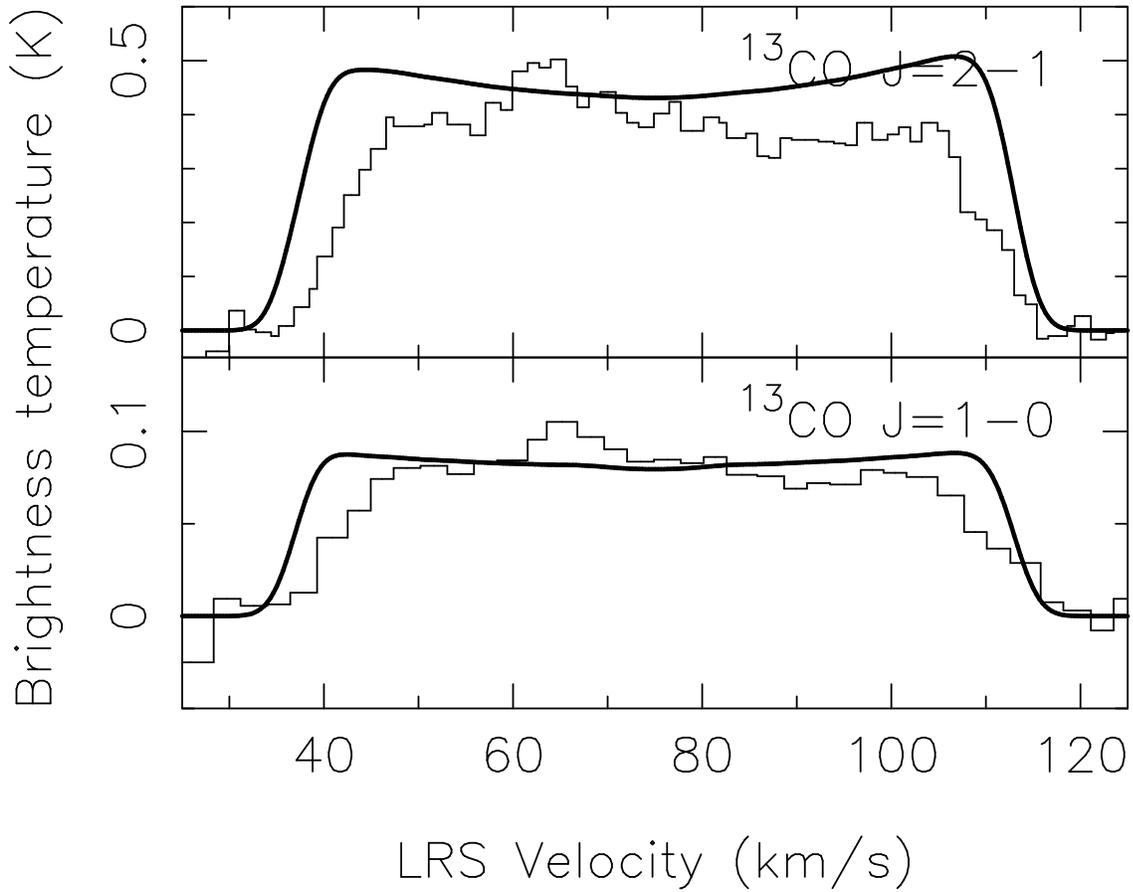}}}}
\end{picture}
\caption{Comparison between the observed $^{13}$CO J=1--0 and J=2--1 lines observed by Bujarrabal et al. (2001) and 
our model predictions (thick solid line). The telescope beams are 22\arcsec\, for J=1--0 and 12\arcsec\, for J=2--1 line, respectively.} 
\label{cocomp}
\end{figure}

\begin{figure}[h] \begin{center}
\setlength{\unitlength}{1cm}
\begin{picture}(15.,20.)
\put(0,12){\resizebox{16cm}{!}{\rotatebox{-90}{\includegraphics*{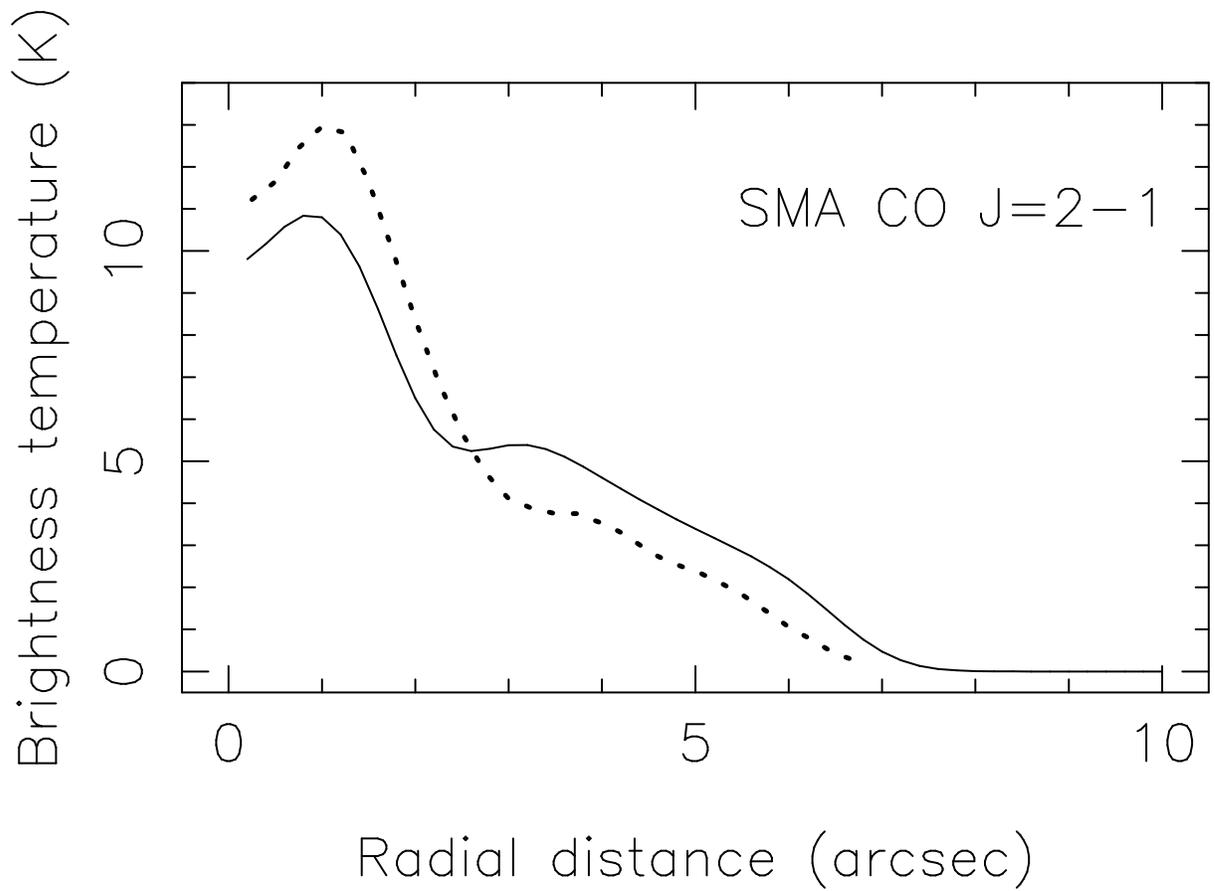}}}}
\end{picture}
\caption{Azimuthal average of the $^{12}$CO J=2--1 brightness temperature (dotted line) obtained from
SMA data at velocity channel V$_{\rm LSR}$ = 75 \kms, and the
model prediction (solid line).}
\label{smacomp}
\end{center} \end{figure}

\begin{figure}[h]
\setlength{\unitlength}{1cm}
\begin{picture}(15.,20.)
\put(0,18){\resizebox{15cm}{!}{\rotatebox{-90}{\includegraphics*{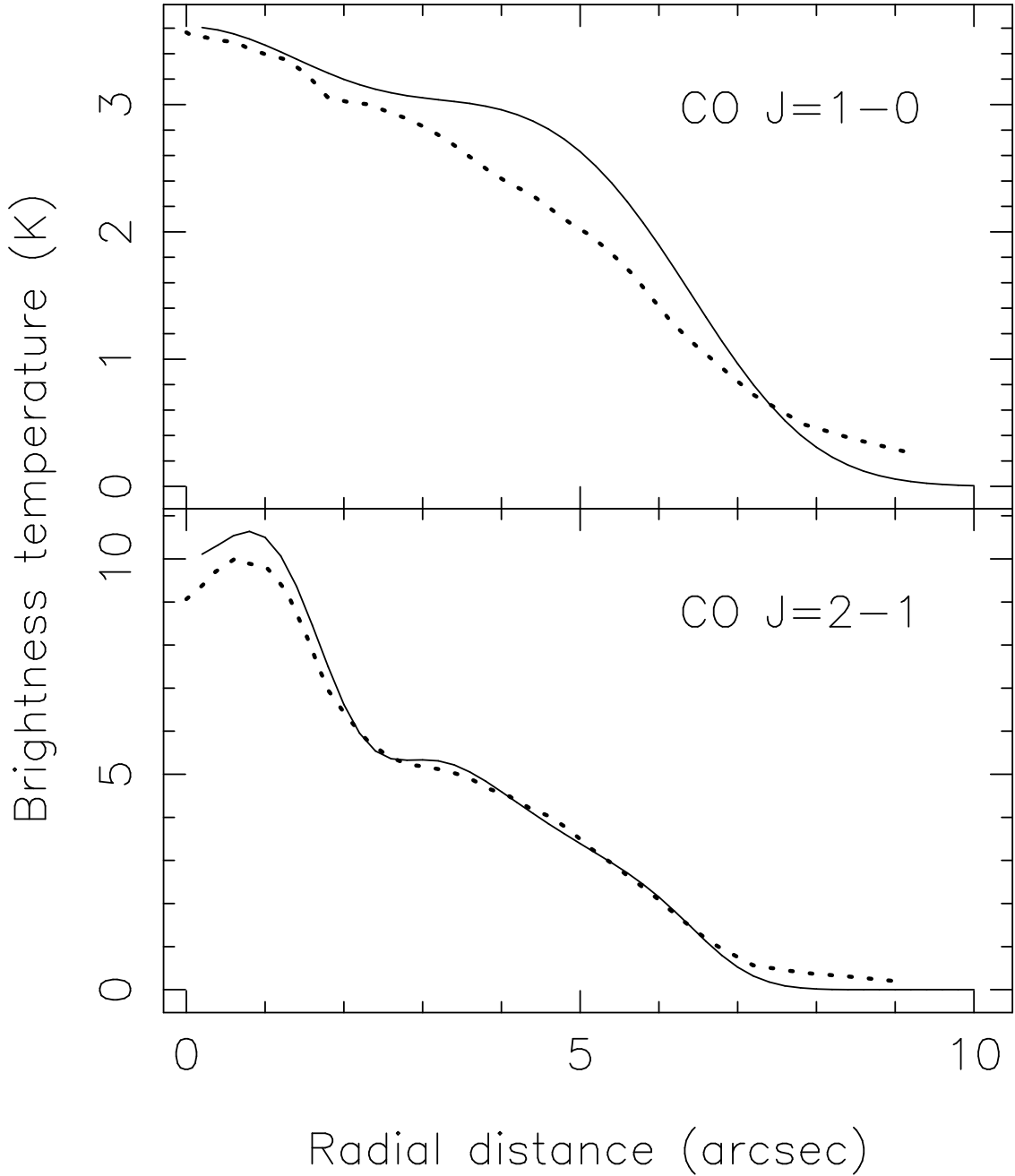}}}}
\end{picture}
\caption{Comparison between the radial distribution of surface brightness (dotted line) of CO J=1--0 and
J=2--1 observed with the Plateau de Bure interferometer (Castro-Carrizo et al. 2007) and our 
model predictions (solid line). The angular resolution is 2\arcsec.86 for the CO J=1--0 and 1\arcsec.4 for the CO J=2--1, 
respectively} 
\label{cocomp}
\end{figure}

\begin{thebibliography}{}
%
\bibitem[Bl\"{o}cker et al. (1999)]{blocker99} Bl\"{o}cker, T., Balega, Y., Hofman, K.H., Lichtenthaler, J., 
Osterbart, R., Weigelt, G., 1999, A\&A 348, 805
\bibitem[Bujarrabal et al. (2001)]{bujarrabal01} Bujarrabal, V., Castro-Carrizo, A., Alcolea, J.,
S\'{a}nchez-Contreras, C., 2001, A\&A, 377, 868
\bibitem[Castro-Carrizo et al. (2007)]{castro07}
Castro-Carrizo, A., Quintana-Lacaci, G., Bujarrabal, V., Neri, R., Alcolea, J., 2007, 465, 457
\bibitem[Castro-Carrizo et al. (2001)]{cas01}
Castro-Carrizo, A., Lucas, R., Bujarrabal, V., Colomer, F., Alcolea, J., 2001, A\&A 368, L34
\bibitem[Davies et al. (2007)]{davies07} Davies, B., Oudmaijer, R.D., Sahu, K.C., 2007, astro-ph/0708.2204
\bibitem[de Jager (1998)]{de98} de Jager, C., 1998, AARev, 8, 145
\bibitem[Dinh-V-Trung \& Nguyen-Q-Rieu (2000)]{trung00}
Dinh-V-Trung, Nguyen-Q-Rieu, 2000, A\&A 361, 601
\bibitem[Flower (2001)]{flower01} Flower, D.R., 2001, JPhB 34, 2731
\bibitem[Goldreich \& Scoville (1976)]{goldreich76}
Goldreich, P., Scoville, N., 1976, ApJ 205, 144
\bibitem[Groenewegen (1994)]{gro94}
Groenewegen, M., 1994, A\&A 290, 544
\bibitem[Harris \& Lambert (1984)]{harris84}
Harris, M.J., Lambert, D.L., 1984, ApJ 281, 739
\bibitem[Hinkle et al. (1976)]{hinkle76}
Hinkle, K.H., Lambert, D.L., Snell, R.L., 1976, ApJ 210, 684
\bibitem[Humphreys et al. (1997)]{humphreys97} Humphreys, R.M., Smith, N., Davidson, K., Jones, T.J., Gehrz, D.,
Mason, C., 1997, AJ, 114, 2778
\bibitem[Jones et al. (1993)]{jones93} Jones, T.J., Humphreys, R.M., Gehrz, R.D., Lawrence, G.F., et al.,
1993, ApJ 411, 323
\bibitem[Justtanont et al. (1994)]{justtanont94}
Justtanont, K., Skinner, C.J., Tielens, A.G.G.M., 1994, ApJ 435, 852
\bibitem[Kastner \& Weintraub (1995)]{kasner95} Kastner, J.H., Weintraub, D.A., 1995, ApJ, 452, 833
\bibitem[Kemper et al. (2003)]{kemper03} Kemper, F., Stark, R., Justtanont, K., de Koter, A., Tielens, A.G.G.M.,
Waters, L.B.F.M., Cami, J., Dijkstra, C., 2003, A\&A, 407, 609
\bibitem[Klochkova et al. (1997)]{klochkova97} Klochkova, V.G., Chentsov, E.L., Panchuk, V.E., 1997, MNRAS 292, 19
\bibitem[Knapp \& Morris (1985)]{knapp85}
Knapp, G.R., Morris, M., 1985, ApJ 292, 640
\bibitem[Mastrodemos et al. (1996)]{Mastrodemos96}
Mastrodemos N., Morris M., Castor J., 1996, ApJ 468, 851
\bibitem[Menten \& Alcolea (1995)]{menten95} Menten, K.M., Alcolea, J., 1995, ApJ 448, 416
\bibitem[Muller et al. (2007)]{muller07} Muller, S., Dinh-V-Trung, Lim, J., Hirano, N., Muthu, C., Kwok, S., 2007, 
ApJ 656, 1109
\bibitem[Nedoluha \& Bowers (1992)]{nedoluha92} Nedoluha, G.E., Bowers, P.F., 1992, ApJ, 392, 249
\bibitem[Omont et al. (1993)]{omon1993} Omont, A., Lucas, R., Morris, M., Guilloteau, S., 1993, A\&A, 267, 490
\bibitem[Oudmaijer et al. (1996)]{oudmaijer96} Oudmaijer, R.D., Groenewegen M.A.T., Matthews, H.E., Blommaert, J.A.D.L.,
Sahu, K.C., 1996, MNRAS, 280, 1062
\bibitem[Oudmaijer (1998)]{Oudmaijer98} Oudmaijer, R.D., 1998, A\&AS 129, 541 
\bibitem[Quintana-Lacaci et al. (2007)]{quintana07} Quintana-Lacaci, G., Bujarrabal, V., Castro-Carrizo, A., Alcolea, J., 2007, A\&A 471, 551
\bibitem[Sahai \& Wannier (1992)]{sahai92} Sahai, R., Wannier, P.G., 1991, ApJ 394, 320
\bibitem[S\'{a}nchez-Contreras et al. (2000)]{sc00} S\'{a}nchez-Contreras, C., Bujarrabal, V., Neri, R., Alcolea, J., 2000,
357, 651
\bibitem[Teyssier et al. (2006)]{teyssier06}
Teyssier, D., Hernandez, R., Bujarrabal, V., Yoshida, H., Phillips, T.G., 2006, A\&A 450, 167
\bibitem[Volk \& Kwok (1988)]{volk88} Volk, K., kwok, S., 1988, ApJ, 331, 435
\bibitem[Walmsley et al. (1991)]{walmsley91} Walsmley, C.M., Chini, R., Kreysa, E., Steppe, H., Forveille, T., 
Omont, A., 1991, A\&A 248, 555
\bibitem[Willacy \& Millar (1997)]{willacy97}
Willacy, K., Millar, T.J., 1997, A\&A 324, 237
%
\end{thebibliography}
\end{document}